      [1999/12/01 v1.4c Il Nuovo Cimento]
\documentclass{cimento}

\usepackage{xspace}
\usepackage{amssymb}

\newcommand{\He}{H.E.S.S.\xspace}
\newcommand{\Ve}{VERITAS\xspace}
\newcommand{\HeII}{H.E.S.S.~II\xspace}
\newcommand{\Ma}{MAGIC\xspace}
\newcommand{\MaII}{MAGIC-II\xspace}

\usepackage{graphicx}  % got figures? uncomment this

\title{EBL studies with ground-based VHE $\gamma$-ray detectors: Current status
and potential of next-generation instruments}

\author{M.~Raue\from{ins:1} \&
D.~Mazin\from{ins:2}}
\instlist{\inst{ins:1}Institut f\"ur Experimentalphysik, Universit\"at Hamburg, Hamburg, Germany
  \inst{ins:2} Institut de Fisica d'Altes Energies (IFAE), Edifici Cn. Universitat Autonoma de Barcelona, 08193 Bellaterra (Barcelona), Spain}
\PACSes{\PACSit{95.85.Pw, 98.70.Vc}{}}
\begin{document}

\maketitle

%------------------------------------------------------------------------------------------------------------------------
% Abstract
%------------------------------------------------------------------------------------------------------------------------
\begin{abstract}
The diffuse meta-galactic radiation field at ultraviolet to infrared wavelengths - commonly labeled extragalactic background light (EBL) - contains the integrated emission history of the universe. Difficult to access via direct observations, indirect constraints on its density can be derived through observations of very-high energy (VHE; E$>$100\,GeV) $\gamma$-rays from distant sources: the VHE photons are attenuated via pair-production with the low energy photons from the EBL, leaving a distinct imprint in the VHE spectra measured on earth. Discoveries made with current-generation VHE observatories like H.E.S.S., MAGIC and VERITAS enabled strong constraints on the density of the EBL, especially in the near-infrared. Here, the constrains on the EBL density from such ground based VHE observations will be briefly reviewed and the potential of the next-generation instruments to improve on these limits will be discussed.
\end{abstract}

%------------------------------------------------------------------------------------------------------------------------
% Introduction
%------------------------------------------------------------------------------------------------------------------------
\section{Introduction}

The observation of very-high energy $\gamma$-rays (VHE; $E>100$\,GeV) from distant sources offers the unique possibility to probe the density of the meta-galactic radiation field at ultraviolet (UV) to infrared (IR) wavelengths, which is commonly labeled the extragalactic background light (EBL; typically 0.1-100\,$\mu$m). The VHE $\gamma$-rays interact with the low energy EBL photons via the pair production process ($\gamma_{\mathrm{VHE}} \gamma_{\mathrm{EBL}} \rightarrow e^+ e^-$) and the flux is attenuated \cite{nikishov:1962a,gould:1967a}. This attenuation can leave distinct signatures in the measured VHE spectra. With assumptions about the source physics and the spectrum emitted at the source location (intrinsic spectrum), constraints on the density of the EBL can be derived (e.g. \cite{stecker:1994a,dwek:1994a}).

The current generation of VHE instruments (\He, \Ma, \Ve) significantly increased the number of known extragalactic VHE sources from 4 in the year 2003 to 44 today (Dec. 2010), most of them being high-frequency peaked BL Lac objects (29), but also intermediate and low frequency peaked BL Lac objects (7), flat spectrum radio quasars (3), radio galaxies (3), and starburst galaxies (2)\footnote{http://tevcat.uchicago.edu/}.  These discoveries, combined with the advanced spectral resolution of these instruments and the wide energy range they cover, led to new strong constraints on the EBL density, in particular at optical to near-IR (NIR)  wavelengths \cite{aharonian:2006:hess:ebl:nature,mazin:2007a,albert:2008:magic:3c279:science}. Since these limits depend on assumptions about the source physics, the strong constraints also sparked intense discussions on the validity of the assumptions and possible caveats (e.g. \cite{aharonian:2002c, katarzynski:2006a, reimer:2007a,aharonian:2008a,krennrich:2008a,raue:2009b}). These discussions have not yet converged and there are interesting arguments for both sides. The operating VHE instruments are being upgraded (\MaII, \HeII). In the future the Cherenkov Telescope Array (CTA\footnote{http://www.cta-observatory.org/}; \cite{hermann:2007a}) will lead to an order of magnitude improvement in sensitivity over current generation experiments covering an extended energy range from 20\,GeV to 100\,TeV.

New possibilities also arise from observations with the Large Area Telescope (LAT; \cite{atwood:2009:fermi:lat:technical}) on board the Fermi satellite. The detector has a large field of view (2.4\,sr) and is operated mostly in survey mode covering the full sky every 3\,h. Fermi/LAT observations span the energy range from 100\,MeV to 300\,GeV and thereby enable measurements of the parts of the energy spectrum, which are not affected by EBL attenuation.

%------------------------------------------------------------------------------------------------------------------------
% The extragalactic background light (EBL)
%------------------------------------------------------------------------------------------------------------------------
\section{The extragalactic background light (EBL): direct measurements}

%------------------
% FIGURE: EBL measurements
\begin{figure}[tbp]
\centering
\includegraphics[width=0.9\textwidth]{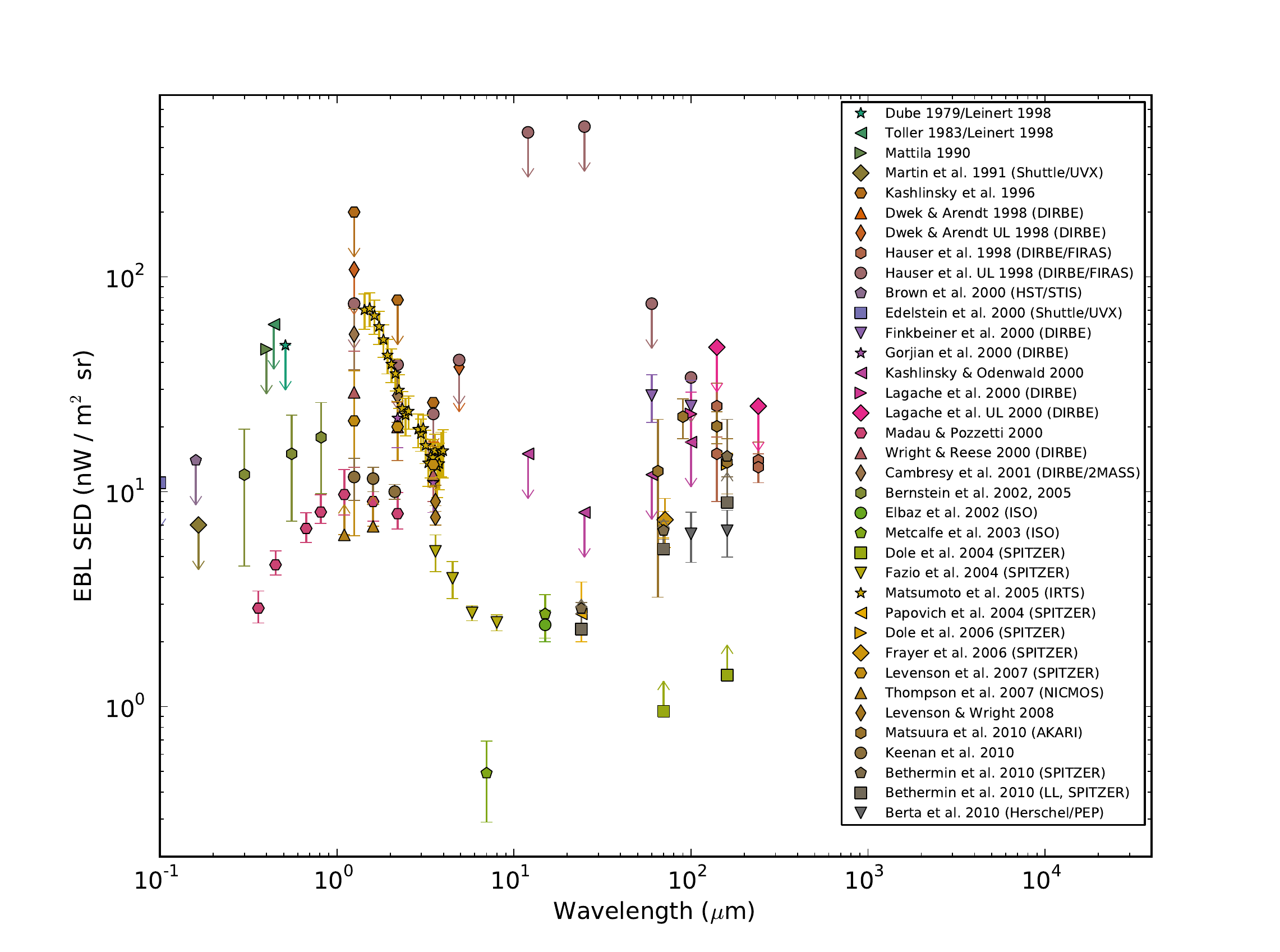}
\caption{Direct measurements and limits on the EBL density (updated version of Fig. 1 from \cite{mazin:2007a} with data from \cite{frayer:2006a, levenson:2007a,thompson:2007a, levenson:2008a, matsuura:2010a, keenan:2010a, bethermin:2010a, berta:2010a}; data compilation and figure available from \texttt{http://www.desy.de/$\sim$mraue/ebl/}).}
\label{Fig:EBLMeasurements}
\end{figure}
 %------------------

Direct measurements of the EBL, especially in the infrared, have proven to be a difficult task due to dominant foregrounds mainly from inter-planetary dust (zodiacal light) \cite{hauser:1998a}. Lower limits on the EBL are derived from integrated source counts (e.g. \cite{madau:2000a,fazio:2004a,frayer:2006a}) and their extrapolation via stacking \cite{dole:2006a}. Recent progress in this field was made with observations by the Herschel instrument (launched in May 2009) and the BLAST mission (flown in 2006) in the FIR resulting in redshift resolved measurements of the EBL  (e.g. \cite{berta:2010a}). Several upper limits were reported from direct observations (e.g. \cite{hauser:1998a}) and from fluctuation analyses (e.g. \cite{kashlinsky:2000a}). The direct limits on the EBL in the UV to far-IR confirm the expected two peak structure although the absolute level of the EBL density remains uncertain by a factor of 2 to 10 (see Fig.~\ref{Fig:EBLMeasurements}).

Several direct detections of the EBL were also reported. They do not contradict the limits but lie significantly above the lower limits derived from integrated resolved sources (galaxies) (see \cite{hauser:2001a} and \cite{Kashlinsky2005:EBLReview} for reviews). In particular, a significant excess of the EBL in the near IR (1 to 4\,$\mu$m), exceeding the expectations from source counts, was reported by the IRTS satellite \cite{matsumoto:2005a}, which initiated a controversial discussion about its origin. If this claimed excess of the EBL is real, it might be attributed to emissions by the first stars in the history of the universe (Population III) and would make the EBL and its structure a unique probe of the epoch of Population III formation and evolution \cite{kashlinsky:2005a}.  The excess, however, overpredicts the number of Ly-$\alpha$ emitters in ultradeep field searches \cite{salvaterra:2006a} and could be an artifact of the subtraction of foreground emissions \cite{dwek:2005c}. It is also strongly disfavored by EBL limits derived from the observations of distant sources of VHE $\gamma$-rays (see next section).

%------------------------------------------------------------------------------------------------------------------------
% EBL limits from VHE observations
%------------------------------------------------------------------------------------------------------------------------
\section{EBL limits from HE and VHE observations}

VHE sources, used to derive limits on the EBL density, belong to a single source class: the active galactic nuclei (AGNs) with the majority of them being blazars (AGNs with strong jet activity and the jets are closely aligned to the line of sight of the observer). Up to now, mainly two different methods - and thereby assumptions about the source intrinsic spectrum - have been utilized to derive limits on the EBL density from VHE observations:
\begin{itemize}
\item \textit{Spectral concavity.} It is assumed that the overall intrinsic source spectrum at high energies will follow a convex shape, or at least will not show an exponential rise towards the highest energies. These assumptions are well motivated by the common leptonic modeling of the sources under investigation (blazars), although different (maybe more exotic models) can possibly reproduce such a feature (e.g. \cite{aharonian:2002c}). Limits on the EBL density are derived by excluding EBL densities that would lead to such features in the observed sources. This method naturally probes the EBL at wavelengths from the mid to the far-infrared.
\item \textit{Maximum spectral hardness.} To probe the EBL in the optical to near-infrared, it is assumed that the intrinsic source spectrum cannot exceed a certain absolute hardness. While somewhat similar in spirit to the first method the underlying assumptions are stronger, since in the energy range of interest (100\,GeV to several TeV) the spectral shape of the intrinsic spectrum is more uncertain. While most of the basic models used to describe the source spectra indeed imply that the VHE spectrum does not exceed a certain hardness, the absolute value is less certain\footnote{e.g., $\Gamma = 1.5 - 0.6$ for $dN/dE \sim E^{-\Gamma}$ and simple leptonic models.} and possible source intrinsic effect (e.g., internal absorption \cite{aharonian:2008a}) could complicate the situation.
\end{itemize}

Strong limits on the EBL density have been derived utilizing the methods sketched above \cite{aharonian:2006:hess:ebl:nature,mazin:2007a,albert:2008:magic:3c279:science}. Especially in the NIR the limits are only a factor $\sim$2 above the lower limits from integrated source count and strongly disfavor an extragalactic origin of the claimed NIR excess (see previous section). In 2010 many new extragalactic VHE sources were discovered, mainly following up on source detected by the Fermi/LAT in the GeV energy range. The spectral properties of many of these sources are not yet known but (i) they cover a large range in redshift (up to $z\sim0.4$), which should give interesting new constrains. (ii) most of them also have data in the HE band (Fermi) and this enables new and interesting methods to derive limits on the EBL density (see next paragraph). (iii) we are approaching a sufficient number of sources ($O(50)$) to begin investigating properties of the population.

Observations made with the Fermi/LAT in the energy range between 100\,MeV and 300\,GeV enable new insights into the spectral properties of VHE blazars:
\begin{enumerate}
\item Fermi/LAT measured the energy spectrum for most of the VHE detected blazars and found their spectra compatible with power laws with spectral indices $\Gamma \gtrsim 1.5$ (e.g.  \cite{abdo:2009:fermi:tevselectedagn,prandini:2010a}). Assuming a low EBL density model (e.g. \cite{franceschini:2008a}) the VHE spectrum for many sources is well described by a simple extrapolation of the Fermi/LAT measured spectrum plus EBL attenuation. A few sources show a softer spectrum at VHE than expected from the Fermi/LAT extrapolation, but no harder spectrum is found. These findings strengthen the assumptions used to derive limits on the EBL from VHE spectra described above.
\item The Fermi/LAT measured spectra of VHE blazars can be used to derive less model dependent limits on the EBL density. Here, it is assumed that the Fermi/LAT measured spectrum is, if extrapolated to VHE, an upper limit to the intrinsic flux from the source. Such methods have been successfully applied to limit the EBL density \cite{georganopoulos:2010a} and to estimate the distance of VHE blazars \cite{prandini:2010a}. \cite{mankuzhiyil:2010a} proposed a method which utilizes the fit of a simple one-zone SSC model to the MWL data at sub-VHE to predict the intrinsic spectrum at VHE. This method enables to measure the absolute EBL attenuation (though in a model dependent approach).
\end{enumerate}

%------------------------------------------------------------------------------------------------------------------------
% Future VHE instruments
%------------------------------------------------------------------------------------------------------------------------
\section{Future VHE instruments}

%------------------
% FIGURE: EBL measurements
\begin{figure}[tbp]
\centering
\includegraphics[width=0.6\textwidth]{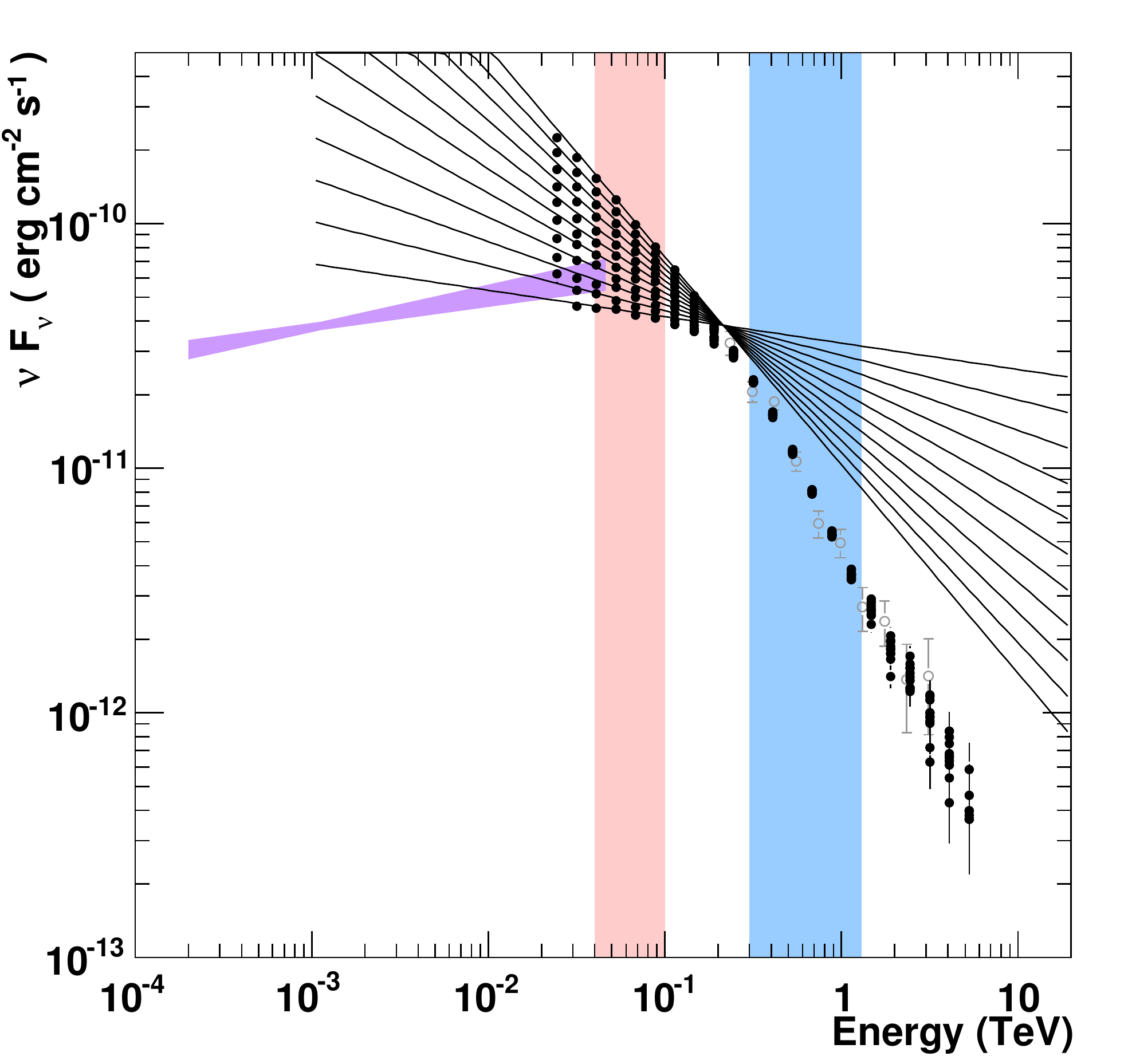}
\caption{Simulated VHE spectrum for PKS\,2155-304 ($z=0.116$) for different assumptions about the EBL density as detected by a NGCTS (see \cite{raue:2010a} for details).}
\label{Fig:PKS2155CTA}
\end{figure}
 %------------------

Current-generation systems have recently been upgraded (\MaII) or the upgrades are under construction (\HeII). These upgrades are mainly aimed to improve the overall sensitivity by a factor two to three and extend the energy range toward the lower energy regime of 20 to 100\,GeV. This will lead to some improvements, but a quantitative difference or a breakthrough compared to the performance of the existing facilities will only be achieved with an order of magnitude improvement in sensitivity. The Next Generation Cherenkov Telescope Systems (NGCTS) are in the advanced planing phase aiming to achieve this order of magnitude improvement: the Cherenkov Telescope Array (CTA\footnote{http://www.cta-observatory.org/} \cite{hermann:2007a}) and the Advanced Gamma-ray Imaging System (AGIS\footnote{http://www.agis-observatory.org/} \cite{buckley:2008a}). CTA envisions to improve the sensitivity over a wide energy range from the few tens of GeV to the multi TeV regime.

Such a wide energy coverage combined with the high sensitivity will enable new studies of the EBL (e.g. \cite{raue:2010a}): (i) the NGCTS will sample parts of the spectrum which are not affected by EBL attenuation. It can be used to derive model-independent results on the EBL density (see Fig.~\ref{Fig:PKS2155CTA} for an example). The advantage of using data from a single instrument over the the combined data from, e.g., Fermi/LAT and current-generation VHE instruments is the homogenous response, the high sensitivity exactly in the transition region between non-attenuated and attenuated regimes, and the reduced systematic uncertainties. (ii) the two peak structure of the EBL is expected to leave an attenuation signature (bump) in the energy range between 1 to 4\,TeV. This is exactly the energy range where a NGCTS will have its highest sensitivity. The detection of a such signature would enable to derive an absolute measurement of the EBL density (not only a limit). (iii) A precision measurement of the cut-off of distant sources at energies above 20\,TeV will enable strong limits on the MIR EBL. (iv) if a NGCTS has  sufficient fluent sensitivity at energies between 20 and 40\,GeV very distant sources with redshift $z>4$ will be in the range of detection. This will be particularly interesting for studies of gamma ray bursts (GRB), which remain so far undetected by the current generation of Cherenkov telescopes.

 \acknowledgments
The authors would like to acknowledge helpful comments from the referee. MR acknowledges financial support from the LEXI program, Hamburg. DM acknowledges the support by a Marie Curie Intra European Fellowship within
the 7th European Community Framework Programme.

%------------------------------------------------------------------------------------------------------------------------
% Bibliography
%------------------------------------------------------------------------------------------------------------------------

\def\Journal#1#2#3#4{{#4}, {#1}, {#2}, #3}
\def\NAT{Nature}
\def\AAA{A\&A}
\def\ApJ{ApJ}
\def\AJ{Astronom. Journal}
\def\Aph{Astropart. Phys.}
\def\ApJS{ApJSS}
\def\MNRAS{MNRAS}
\def\NIM{Nucl. Instrum. Methods}
\def\NIMA{Nucl. Instrum. Methods A}
% Bibliography and bibfile
\def\aj{AJ}%
          % Astronomical Journal
\def\actaa{Acta Astron.}%
          % Acta Astronomica
\def\araa{ARA\&A}%
          % Annual Review of Astron and Astrophys
\def\apj{ApJ}%
          % Astrophysical Journal
\def\apjl{ApJ}%
          % Astrophysical Journal, Letters
\def\apjs{ApJS}%
          % Astrophysical Journal, Supplement
\def\ao{Appl.~Opt.}%
          % Applied Optics
\def\apss{Ap\&SS}%
          % Astrophysics and Space Science
\def\aap{A\&A}%
          % Astronomy and Astrophysics
\def\aapr{A\&A~Rev.}%
          % Astronomy and Astrophysics Reviews
\def\aaps{A\&AS}%
          % Astronomy and Astrophysics, Supplement
\def\azh{AZh}%
          % Astronomicheskii Zhurnal
\def\baas{BAAS}%
          % Bulletin of the AAS
\def\bac{Bull. astr. Inst. Czechosl.}%
          % Bulletin of the Astronomical Institutes of Czechoslovakia 
\def\caa{Chinese Astron. Astrophys.}%
          % Chinese Astronomy and Astrophysics
\def\cjaa{Chinese J. Astron. Astrophys.}%
          % Chinese Journal of Astronomy and Astrophysics
\def\icarus{Icarus}%
          % Icarus
\def\jcap{J. Cosmology Astropart. Phys.}%
          % Journal of Cosmology and Astroparticle Physics
\def\jrasc{JRASC}%
          % Journal of the RAS of Canada
\def\mnras{MNRAS}%
          % Monthly Notices of the RAS
\def\memras{MmRAS}%
          % Memoirs of the RAS
\def\na{New A}%
          % New Astronomy
\def\nar{New A Rev.}%
          % New Astronomy Review
\def\pasa{PASA}%
          % Publications of the Astron. Soc. of Australia
\def\pra{Phys.~Rev.~A}%
          % Physical Review A: General Physics
\def\prb{Phys.~Rev.~B}%
          % Physical Review B: Solid State
\def\prc{Phys.~Rev.~C}%
          % Physical Review C
\def\prd{Phys.~Rev.~D}%
          % Physical Review D
\def\pre{Phys.~Rev.~E}%
          % Physical Review E
\def\prl{Phys.~Rev.~Lett.}%
          % Physical Review Letters
\def\pasp{PASP}%
          % Publications of the ASP
\def\pasj{PASJ}%
          % Publications of the ASJ
\def\qjras{QJRAS}%
          % Quarterly Journal of the RAS
\def\rmxaa{Rev. Mexicana Astron. Astrofis.}%
          % Revista Mexicana de Astronomia y Astrofisica
\def\skytel{S\&T}%
          % Sky and Telescope
\def\solphys{Sol.~Phys.}%
          % Solar Physics
\def\sovast{Soviet~Ast.}%
          % Soviet Astronomy
\def\ssr{Space~Sci.~Rev.}%
          % Space Science Reviews
\def\zap{ZAp}%
          % Zeitschrift fuer Astrophysik
\def\nat{Nature}%
          % Nature
\def\iaucirc{IAU~Circ.}%
          % IAU Cirulars
\def\aplett{Astrophys.~Lett.}%
          % Astrophysics Letters
\def\apspr{Astrophys.~Space~Phys.~Res.}%
          % Astrophysics Space Physics Research
\def\bain{Bull.~Astron.~Inst.~Netherlands}%
          % Bulletin Astronomical Institute of the Netherlands
\def\fcp{Fund.~Cosmic~Phys.}%
          % Fundamental Cosmic Physics
\def\gca{Geochim.~Cosmochim.~Acta}%
          % Geochimica Cosmochimica Acta
\def\grl{Geophys.~Res.~Lett.}%
          % Geophysics Research Letters
\def\jcp{J.~Chem.~Phys.}%
          % Journal of Chemical Physics
\def\jgr{J.~Geophys.~Res.}%
          % Journal of Geophysics Research
\def\jqsrt{J.~Quant.~Spec.~Radiat.~Transf.}%
          % Journal of Quantitiative Spectroscopy and Radiative Trasfer
\def\memsai{Mem.~Soc.~Astron.~Italiana}%
          % Mem. Societa Astronomica Italiana
\def\nphysa{Nucl.~Phys.~A}%
          % Nuclear Physics A
\def\physrep{Phys.~Rep.}%
          % Physics Reports
\def\physscr{Phys.~Scr}%
          % Physica Scripta
\def\planss{Planet.~Space~Sci.}%
          % Planetary Space Science
\def\procspie{Proc.~SPIE}%
          % Proceedings of the SPIE
          
%%\input{../../../../bibtex/journalNames.tex}

% local
%\bibliographystyle{varenna-bibtex/varenna}
%\bibliography{../../../../bibtex/bibtex_db_2011}

\end{document}